\documentclass[conference]{IEEEtran}
\usepackage{amsfonts}
\usepackage{amsmath}
\usepackage{graphics}
\usepackage{url}
\usepackage[pdftex]{graphicx}

\ifCLASSINFOpdf
\else
\fi
\begin{document}

\title{Community Detection in Multiplex Networks using Locally Adaptive Random Walks}
\author{\IEEEauthorblockN{Zhana Kuncheva} 
\IEEEauthorblockA{Department of Mathematics \\  
Imperial College London 
} \and \IEEEauthorblockN{Giovanni Montana} 
\IEEEauthorblockA{Department of Biomedical Engineering \\ 
King's College London}}
\maketitle

\begin{abstract}
\boldmath
Multiplex networks, a special type of multilayer networks, are
increasingly applied in many domains ranging from social media analytics to
biology. A common task in these applications concerns the detection of
community structures. Many existing algorithms for community detection in
multiplexes attempt to detect communities which are shared by all
layers. In this article we propose a community detection algorithm, LART (Locally Adaptive Random Transitions), for the detection of communities that are shared by either some or all the layers in the multiplex. The algorithm is based on a random walk on the multiplex, and the transition probabilities defining the random walk
are allowed to depend on the local topological similarity between layers at
any given node so as to facilitate the exploration of communities across
layers. Based on this random walk, a node dissimilarity measure is derived
and nodes are clustered based on this distance in a hierarchical fashion. We
present experimental results using networks simulated under various
scenarios to showcase the performance of LART in comparison to related
community detection algorithms.
\end{abstract}

\IEEEpeerreviewmaketitle

\section{Introduction\label{intro}}

Many real world systems, including social and biological ones, are often
represented as complex networks capturing the interacting nature of multiple
agents populating the system~\cite{Strogatz2001}. The different agents are
interpreted as the nodes of the network, and the relations among them are
encoded by the edges of the network. An important aspect of network analysis
is the discovery of community structures defined as groups of nodes that are
more densely connected to each other than they are to the rest of the
network~\cite{Girvan2002}. A large body of work exists on community
detection, and one such extensive review of the area is given by~\cite%
{Fortunato2010}. One useful way to explore the structural properties of a real network is to study the behavior of a discrete-time random walk on it~\cite{Lovasz}. Random walks have
been successfully used to unfold the community structure on a network; see,
for instance,~\cite{Pons},~\cite{zhou2004} and~\cite{Rosvall2008}. The
main intuition behind these approaches is that a random walker that jumps
from node to node with preset transition probabilities is expected to get
``trapped" for longer times in denser regions defining the communities. 

A multiplex is a particular type of a multilayer network where all layers
share the same set of nodes but may have very different topology~\cite%
{Boccaletti2014}. The structure of the multiplex allows for layers to be
connected by inter-layer weights. These weights represent
some type of association between layers, and can be specified either by
using information extracted directly from the multiplex or external
data. There is a wide class of real networks that can be represented by a
multiplex. Some examples are the social interactions between users with respect to various social media or the different transportation means between stations in a city; for a survey see~\cite{Kivel}. One important area of research on multiplex networks is community detection since it can identify shared structures of nodes in the multiple layers. 

In this paper we investigate the task of detecting communities that may be shared by a subset of the layers in the multiplex.
Many multiplex community detection approaches identify a community partition that best fits all given layers, i.e. they detect communities shared by all layers. Some of these methods collapse the
information into a single layer and then use traditional community
detection algorithms for networks,~\cite{Berlinger}, while other methods extend community detection algorithms from one to multiple layers,~\cite{Tang2009a,Amelio2014}. There exist real-world systems, however, for which some communities may be shared only by a subset of layers. Take genomic data as an example, where groups of genes can be associated with specific functional processes relevant to some tissues but not to others~\cite{Dobrin}. Relatively few solutions exist to address this problem of detecting communities in a subset of layers,~\cite{Mucha,DeDomenico2015}, therefore we propose a new approach to tackle the issue. 

The methodology we introduce in this paper is based on a discrete-time random
walk on the multiplex. A multiplex random walk explores the network both within and between layers according to some preset transition probabilities~\cite{DeDome}. We offer a novel approach of adapting these transition probabilities to depend on the local topological similarity between any pair of layers, at any given node. By encouraging jumps between nodes in
different layers sharing similar topology, and penalizing jumps involving
nodes across layers that do not share local topology, we aim to facilitate
the exploration of potential communities that may be shared across layers.
The resulting algorithm, called LART (Locally Adaptive Random Transitions), defines a
multiplex random walker that will spend a longer time moving between nodes in
communities which are shared across layers. The random walk will also get ``trapped" in one
layer for longer if a community is specific to this layer. We take advantage of
the properties of this random walk to introduce a distance measure between
nodes, and an agglomerative clustering procedure is then used to detect communities within and between layers. The resulting algorithm
can be considered as an extension of the WalkTrap algorithm~\cite{Pons} to the multiplex framework.

The paper is organized as follows. In Section~\ref{review}, we provide a
concise literature review of multiplex community detection methods. Section~%
\ref{Sreview} introduces the LART algorithm. In Section~\ref{Example} we provide 
an illustrative example to distinguish between communities shared by two or more layers (shared), and communities
that are specific to one layer (non-shared). In Section~\ref{results} we
compare the performance of the LART algorithm and other multiplex community
detection methods. Our experimental results are based on networks simulated
under various scenarios for illustrative purposes. In Section~\ref{conc} we
provide concluding remarks and directions for future work.

\section{Related Work\label{review}}

In recent years a handful of algorithms have been described in the
literature to address the problem of finding community structures that are
shared by all layers. A straightforward approach relies on a simple layer aggregation
procedure whereby all the layers are first collapsed into a single network so that traditional algorithms for community detection can be used afterward. The weights of the edges between any two nodes in the aggregated network are defined as a linear combination of the weights between those same nodes from each of the layers~\cite{Berlinger}, and different assignments of these weights have been discussed in \cite{Berlinger, Berlingerio2011, Zhu2014, Cai2005}. Another direction
consists of applying a community detection algorithm to each separate layer,
and then combining all the resulting partitions either by using cluster ensemble
approaches~\cite{Strehl2000} or by merging communities across all layers such that each multiplex community contains a predefined minimum number of corresponding nodes~\cite{Berlingerio2013}.

Extensions of community detection algorithms from one to
multiple layers have also been proposed in an attempt to take into account as
much information as possible for each layer. Two such
examples rely on the extension of a function of modularity $Q$~\cite%
{Girvan2002}. In its original form, the modularity $Q$ is defined as the
number of connections within a community compared to the expected number of
such connections in an equivalent random network. In~\cite{Tang2009a}, Principal Modularity Maximization concatenates the partitions
obtained on each separate layer by maximizing modularity $Q$. Using results
from Generalized Canonical Correlation Analysis, the authors obtain the
final partition by computing the top $k$ eigenvectors of the concatenated
matrix. Another extension of modularity $Q$ for a multiplex network is
proposed in~\cite{Mucha}. In addition to the usual interpretation of $Q$, this extension accounts for inter-layer weights that exist
between nodes in different layers. In this way, communities that already
exist in separate layers can be coupled. The two methods in~\cite{Tang2009a} and~\cite{Mucha} are robust to
noise and variation in the layers.

The authors of \cite{DeDomenico2015} provide an extension to a flow-based and information-theoretic algorithm known as Infomap~\cite{Rosvall2008}. The information flow is modeled as a random walk with teleportation through the network. The best community partition over the network is scored by minimizing the map equation, which measures the description length of the random walker within and between communities. In \cite{DeDomenico2015} Infomap is generalized for multiplex networks by modifying some of the constraints in the original map equation to allow for nodes in different layers to be assigned to different communities. 

Another method for multiplex community detection is the top-bottom network
partitioning approach~\cite{Brodka}, which uses a cross-layer
edge clustering coefficients to decide whether communities should be split.
In~\cite{Amelio2014}, a multi-objective optimization algorithm iteratively
maximizes the modularity of the current layer while simultaneously
maximizing the similarity between the current and the previous layers. The
authors of \cite{Hmimida} extend a seed-centric algorithm to fit the
multiplex network by means of multiplex centrality measures. Following their work
on subspace clustering methods used to detect a set of relevant layers for each community,
the authors of \cite{Boden2012} use a search tree for detecting communities. Starting from a seed node, the proposed algorithm iteratively expands the communities with respect to a quality function that can be specified by the user.

Since a multiplex can be represented as a third order tensor \cite{DeDomenico2013a}, tensor
decompositions have also been investigated for the problem of community
detection in multiple layers; some of many such examples are~\cite{Ng2014}, \cite{Papalexakis2013} and \cite{Gauvin2014}. These methods
obtain the partitioning using different tensor factorizations. They are
advantageous since they are fast due to their closed-form solution, although the number of
communities needs to be specified in advance.

\section{Methods\label{Sreview}}

In this section, we first discuss how the inter-layer weights at each node can be defined using a topological similarity measure, and introduce a
supra-adjacency matrix in which within-layer and inter-layer connections are
stored. We then adapt the transition probabilities of a multiplex random walk to depend on the local topological similarity between any pair of layers at any given node, and briefly discuss their properties. In order to group nodes into communities using the
multiplex random walk, we introduce a dissimilarity measure between nodes
that captures the community structure of the multiplex. This measure
considers two separate cases: when two nodes from the same
layer are compared, and when two nodes from
different layers are compared. Lastly we explain how these distances are used to
generate hierarchical clusters and detect communities.

\subsection{The supra-adjacency matrix\label{sa2}}

An $L$-layered multiplex network is a multilayer undirected graph $\mathcal{%
M}=(V;A_{k})_{k=1}^{L}$, where $V$ is a set of nodes, where $|V|=N$, and $A_{k}$ is the $%
N\times N$ adjacency matrix representing the set of edges in layer $L_{k}$
for $k=1,2,...,L$. For any node $v_{i}\in V$, $i=1,2,...,N$, we denote node $%
v_{i}$ in layer $L_{k}$ by $v_{i}^{k}$. The connection between nodes $v_{i}$
and $v_{j}$ in $L_{k}$ is given by $A_{ij;k}=A_{ji;k}$. Nodes $v_{i}$ and $%
v_{j}$ in $L_{k}$ are neighbors if $A_{ij;k}=A_{ji;k}=1$, otherwise $%
A_{ij;k}=0$. Furthermore, $\forall k$, $A_{ij;k}=0$ for $i=j$. The weighted
edge between nodes $v_{i}^{k}$ and $v_{i}^{l}$ is the inter-layer connection
denoted by $\omega _{i;kl}\in \mathbb{R}$. Inter-layer weights have been
modeled in different ways in the existing literature, and usually a fixed
inter-layer value $\omega _{i;kl}=\omega$ $%
\forall i,k,l,$ is adopted. Different values of $\omega \in [0,\infty)$ have been considered to analyze how they affect the time required for a random walk to cover all nodes in the multiplex,~\cite{DeDome,Sole-Ribalta2013}. In~\cite{Mucha}, $\omega \in (0,1]$ is
interpreted as uniform coupling strength. Values of $\omega$ close to $1$ encourage 
the same community assignment of a node in two different layers, while $\omega$ close to $0$ 
does not support coupling of communities from different layers. 

In this work, the inter-layer weights $\omega _{i;kl}$ reflect the
similarity in local topology between $v_{i}^k$ and $v_{i}^l$, and is defined
as the number of edges that the two nodes have in common between layers,
i.e. 
\begin{equation*}
\omega _{i;kl}:=|N_{i,k}\cap N_{i,l}| 
\end{equation*}
where $N_{i,k}:=\{v_{j}^k:A_{ij;k}=1\} $ is the set of edges for $v_{i}^k$.
It follows that $\omega _{i;kl}\in[0,N-1]$.


The $\mathcal{M}$ network has an associated $(NL\times NL)$ block matrix
called the supra-adjacency matrix $\mathcal{A}^{*}$. The diagonal blocks are
the adjacency matrices $A_{k}$, and the off-diagonal blocks are the
inter-layer connection diagonal matrices $W_{km}$, namely $%
W_{km}=diag(\omega _{1;km},\omega _{2;km},...,\omega _{N;km})$. Thus $%
\mathcal{A}_{(i,k)(j,l)}^{*}$ indicates the connection between node $%
v_{i}^{k}$ and node $v_{j}^{l}$.

The $\mathcal{A}^{\ast}$ matrix is used to define the multiplex random walk
on $\mathcal{M}$. In order to define a random walk that is
well-suited for exploring the whole multiplex, however, we require $\mathcal{%
A}^{\ast }$ to be \textquotedblleft well-behaved", i.e. connected and
non-bipartite~\cite{Lovasz}. A network is connected if there exists a path
between any two pairs of nodes, and is bipartite if it can be divided
into two disjoint sets such that no links connect two nodes in the same set. 
$\mathcal{A}^{\ast }$ is not necessarily connected since we may have
inter-layer connections $\omega _{i;kl}=0$. Since $A_{ii;k}=0$ for $\forall
i,k$, $\mathcal{A}^{\ast }$ may be also bipartite. We introduce a new
supra-adjacency matrix $\mathcal{A}$ obtained from $\mathcal{A}^{\ast }$ by
replacing the entry $A_{j}$ with $A_{j}+\varepsilon I$ and $W_{ij}$ with $%
W_{ij}+\varepsilon I;$ here $I$ is the $N\times N$ identity matrix and $%
0<\varepsilon \leq 1$. The positive weights on the main diagonal of $%
\mathcal{A}$ make the multiplex non-bipartite, while the positive weights in
the off-diagonal blocks' main diagonals make the multiplex connected. Both $%
\mathcal{A}$ and $\mathcal{A}^{\ast }$ clearly have the same topology. We
use $\mathcal{A}$ to define the transition probabilities in the
next section.  

\subsection{Locally Adapted Random Transition Probabilities\label{sA}}

A discrete-time random walk on $\mathcal{M}$ should be allowed to move
within and across layers. The structure of $\mathcal{M}$ allows four
possible moves that a random walker can make when in node $v_{i}^{k}$:
when it stays in the same layer $L^{k}$, it can either stay at $v_{i}^{k}$ or move to
a neighboring node $v_{j}^{k}$; when it jumps to another layer $L^{l}$, it can
either make a step to its corresponding node, $v_{i}^{l}$, or move to a different one, $%
v_{j}^{l}$. The corresponding transition probabilities associated to these
four possible moves are defined as 
\begin{equation}
\begin{split}
\mathcal{P}_{(i,k)(i,k)}:=& \frac{\mathcal{A}_{(i,k)(i,k)}}{\kappa _{i,k}} \\
\mathcal{P}_{(i,k)(i,l)}:=& \frac{\mathcal{A}_{(i,k)(i,l)}}{\kappa _{i,k}}
\end{split}%
\qquad 
\begin{split}
\mathcal{P}_{(i,k)(j,k)}:=& \frac{\mathcal{A}_{(i,k)(j,k)}}{\kappa _{i,k}} \\
\mathcal{P}_{(i,k)(j,l)}:=& 0
\end{split}
\label{eq:probs}
\end{equation}%
where $\kappa _{i,k}$ is the multiplex degree of node $v_{i}^{k}$ in $%
\mathcal{A}$ defined as $\kappa _{i,k}:=\sum_{j,l}\mathcal{A}_{(i,k)(j,l)}$.
In this formulation, the transition probabilities depend on the topological similarity between
layers. 

The rationale for these definitions is as follows. When $v_{i}^{k}$ and $%
v_{i}^{l}$ have several common neighbors, it may be possible that both nodes
belong to a community shared by those two layers, $L_{k}$ and $L_{l}$; in
this case $\omega _{i;kl}$ is high, and in turn $\mathcal{P}_{(i,k)(i,l)}$
is also high in order to encourage this type of move. In the extreme
case when the local topology of $v_{i}$ is exactly the same across all $L$
layers, then   
\begin{equation*}
\sum_{l=1;l\neq k}^{L}\mathcal{P}_{(i,k)(i,l)}=(L-1)/L
\end{equation*}%
and $\sum_{j}\mathcal{P}_{(i,k)(j,k)}=1/L$. In this setting, for $L=2$, the
random walker will be equally likely to stay at the current layer $L_{k}$ or
explore the other layer; for $L>2$, the random walker will have higher
probability to move to $L_{l}$, $l\neq k$, rather than staying in the current layer $L_{k}$. On the contrary, when a node $%
v_{i}^{k}$ belongs to a community which is specific only to $L_{k}$, we
expect that $\omega _{i;kl}$, $l\neq k$ will be small; in this case, $%
\sum_{l\neq k}\mathcal{P}_{(i,k)(i,l)}\approx 0$, and the random walker will
remain \textquotedblleft trapped"' for longer time in the region of the community on the current layer $%
L_{k}$ since $\sum_{j}\mathcal{P}_{(i,k)(j,k)}\approx 1$. When a node $%
v_{i}^{k}$ is disconnected in $L_{k}$, the random walker will either stay in
the current layer with probability proportional to $\mathcal{A}%
_{(i,k)(i,k)}=\varepsilon $ or move to any other $L_{l}$, $l\neq k$ with
probability proportional to $\sum_{l}\mathcal{A}_{(i,k)(i,l)}=\varepsilon
(L-1)$. 

We note that in a multiplex there exist inter-layer connections only between corresponding nodes. Therefore the probability to move from
a node $v_{i}^{k}$ to other nodes $v_{j}^{l}$ for $i\neq j$ and $k\neq l$ is zero since there cannot exist a direct move where there is no connection.

Our transition probabilities in~\eqref{eq:probs} may be
represented as an $NL\times NL$ transition matrix $\mathcal{P}$ of the
random walk process, which may also be written as   
\begin{equation*}
\mathcal{P}:=\mathcal{D}^{-1}\mathcal{A}
\end{equation*}%
This representation is useful for showing the stationary properties of the random walk. Here $\mathcal{D}$ is the $NL\times NL$ diagonal
matrix defined by the multiplex node degrees, i.e. $\mathcal{D}%
_{(i,k)(i,k)}:=\kappa _{i,k}$ and $\mathcal{D}_{(i,k)(j,l)}=0$ for $i\neq j$
or $k\neq l$. The resulting transition matrix $\mathcal{P}$ is stochastic
since $0\leq \mathcal{P}_{(i,k)(j,l)}\leq 1,\forall i,j,k,l,$ and $%
\sum_{(j,l)}\mathcal{P}_{(i,k)(j,l)}=1$ for every $(i,k)$. Let
the probability distribution 
\begin{equation*}
p(t)=\left[ p_{(i,k)}(t)\right] _{i=1;k=1}^{N,L}
\end{equation*}%
be the vector of probability values for all nodes in the network, where $p_{(i,k)}(t)$ is the probability of finding
the random walker in node $v_{i}^{k}$ after $t$ steps. The dynamics of the
probability distribution $p(t)$ is given by: 
\begin{equation}
p(t+1)=p(t)\mathcal{P}=p(0)\mathcal{P}^{t}.  \label{eq:2}
\end{equation}%
It follows that the probability to start in node $v_{i}^{k}$ and reach node $%
v_{j}^{l}$ in $t$ steps is given by $\mathcal{P}_{(i,k),(j,l)}^{t}$. A
stationary distribution  of $\mathcal{P}$ satisfies the equation: 
\begin{equation}
p^{\ast }=p^{\ast }\mathcal{P},  \label{eq:stat1}
\end{equation}%
with $\sum_{(i,k)}p_{(i,k)}^{\ast }=1$ and $0\leq p_{(i,k)}^{\ast }\leq 1$
for all $(i,k)$, see \cite{Lovasz}. It can be proved that $\mathcal{P}$ is
irreducible and aperiodic since it is defined by $\mathcal{A}$, which is
connected and non-bipartite. Therefore the existence and uniqueness of the
stationary distribution $p^{\ast}$ is guaranteed by the Perron-Frobenius
Theorem~\cite{Lovasz}. The stationary distribution corresponds to the left-eigenvector of $\mathcal{P}$
associated with eigenvalue $1$, and is obtained as 
\begin{equation*}
p_{(i,k)}^{\ast }=\frac{\kappa _{i,k}}{\sum_{j,l}\kappa _{(j,l)}}.
\end{equation*}

There are two implications of the stationary distribution, which we need to consider 
in order to introduce a dissimilarity measure between nodes in the multiplex. First, as the number of steps $t$ tends to
infinity, the probability of being on a node $v_{j}^{l}$ depends only on the
degree of the node $v_{j}^{l}$ regardless of what the starting node is,
i.e.: 
\begin{equation}
\quad \lim_{t\rightarrow \infty }\mathcal{P}_{(i,k)(j,l)}^{t}\rightarrow
p_{(j,l)}^{\ast }=\frac{\kappa _{j,l}}{\sum_{h,m}\kappa _{h,m}},\quad
\forall (i,k).  \label{eq:stat}
\end{equation}

Second, the stationary distribution satisfies the time-reversibility
property of the chain: 
\begin{equation}
\kappa _{i,k}\mathcal{P}_{(i,k)(j,l)}^{t}=\kappa _{j,l}\mathcal{P}%
_{(j,l)(i,k)}^{t},\quad \forall (i,k),\forall (j,l).  \label{eq:prop}
\end{equation}%
These are standard results; see~\cite{Lovasz}. 

The convergence to the stationary distribution implies that for large $t$
all rows of the matrix $\mathcal{P}^{t}$ approach the stationary
distribution~\eqref{eq:stat}, thus the whole multiplex becomes one
community. On the other hand, for smaller $t$ the random walk captures local
community structures (a fact also observed in~\cite{Pons} and~\cite{Rosvall2008}), therefore we use short random walks to detect communities as desired.

The time-reversibility property~\eqref{eq:prop} implies that even if the probability to reach $v_{j}^{l}$ in $t$ steps starting from $v_{i}^{k}$ is high, it does not follow that the probability to reach $v_{i}^{k}$ in $t$ steps starting from $v_{j}^{l}$ is high. Therefore, it is insufficient to compare nodes $v_{i}^{k}$ and $v_{j}^{l}$ only through $\mathcal{P}_{(i,k)(j,l)}^{t}$ or $\mathcal{P}_{(j,l)(i,k)}^{t}$. For this reason we use the $NL$-dimensional vector of probabilities 
\[
\mathcal{P}_{(i,k)(\cdot ,\cdot )}^{t}=\left[ \mathcal{P}_{(i,k)(h,m)}^{t}%
\right] _{m=1,h=1}^{L,N}\] available for a node $v_{i}^{k}$ to define its dissimilarity to any other node. 

\subsection{Node dissimilarity matrix}

In this section we introduce an $NL\times NL$ node dissimilarity matrix $S(t)
$ which depends on the multiplex random walk of length $t$. This
matrix contains all possible distances between any pair of nodes, both
within and between layers. These distances are defined such that, when two
nodes belong to the same community, their distance is low, regardless of
whether they are in the same layer or not; conversely, the distance between
two nodes is large when they are not in the same community, again regardless of the layers they are in.

In order to define the elements of the dissimilarity matrix $S$ we need to
consider two separate cases: in the case when we compare two nodes in
the same layer, they are in the same community if they "see" in a similar way the rest of the nodes
in the current layer and all nodes in the other layers. 

Another case is when we compare two nodes in two separate layers  $L_{k}$
and $L_{l}$;  they will be in the same community only if such a community is
shared by both layers. Therefore, they will ``see" in a
similar way all nodes both in their respective layers $L_{k}$ and $L_{l}$ and in each other's layers $L_{l}$ and $L_{k}$. 

\subsubsection*{Same layer} when $v_{i}^{k}$ and $v_{j}^{k}$ are in the same layer, their
dissimilarity is defined as: 
\begin{align}
S(t)_{(i,k)(j,k)}& :=\sqrt{\sum_{h=1}^{N}\sum_{m=1}^{L}\frac{\left( \mathcal{%
P}_{(i,k)(h,m)}^{t}-\mathcal{P}_{(j,k)(h,m)}^{t}\right) ^{2}}{\kappa _{(h,m)}%
}}=  \notag  \label{eq:def1} \\
& =\left\Vert \mathcal{D}^{-\frac{1}{2}}\mathcal{P}_{(i,k)(.,.)}^{t}-%
\mathcal{D}^{-\frac{1}{2}}\mathcal{P}_{(j,k)(.,.)}^{t}\right\Vert 
\end{align}%
where $\left\Vert .\right\Vert $ is the Euclidean norm. The distance $%
S(t)_{(i,k)(j,k)}$ is small when two nodes from the same layer, $v_{i}^{k}$
and $v_{j}^{k}$, are in the same community, since the probabilities to reach
any other node in layer $L_{k}$ starting from $v_{i}^{k}$ or $v_{j}^{k}$
will be approximately equal, $\mathcal{P}_{(i,k)(h,k)}^{t}\simeq \mathcal{P}%
_{(j,k)(h,k)}^{t},\forall h=1,2,...,N$. Moreover, the probabilities to reach
any other node in any other layer $L_{l}$, $l\neq k$, starting from $%
v_{i}^{k}$ or $v_{j}^{k}$ will be also almost equal, $\mathcal{P}%
_{(i,k)(h,l)}^{t}\simeq \mathcal{P}_{(j,k)(h,l)}^{t},\forall
h=1,2,...,N,\forall l\neq k$.

\subsubsection*{Different layers} When $v_{i}^{k}$ and $v_{j}^{l}$ are in two different layers, $L_{k}$ and 
$L_{l}$, we define the dissimilarity as: 
\begin{equation*}
S(t)_{(i,k)(j,l)}:=\sqrt{s_{1}+s_{2}+s_{3}}
\end{equation*}%
where \textbf{\ 
} 
\begin{align}
s_{1}& :=\sum_{h=1}^{N}\left( \frac{\mathcal{P}_{(i,k)(h,k)}^{t}}{\sqrt{%
\kappa _{(h,k)}}}-\frac{\mathcal{P}_{(j,l)(h,l)}^{t}}{\sqrt{\kappa _{(h,l)}}}%
\right) ^{2}  \notag \\
s_{2}& :=\sum_{h=1}^{N}\left( \frac{\mathcal{P}_{(i,k)(h,l)}^{t}}{\sqrt{%
\kappa _{(h,l)}}}-\frac{\mathcal{P}_{(j,l)(h,k)}^{t}}{\sqrt{\kappa _{(h,k)}}}%
\right) ^{2}  \notag \\
s_{3}& :=\sum_{h=1}^{N}\sum_{\substack{ m=1; \\ m\neq k,l}}^{L}\frac{\left( 
\mathcal{P}_{(i,k)(h,m)}^{t}-\mathcal{P}_{(j,l)(h,m)}^{t}\right) ^{2}}{%
\kappa _{(h,m)}}.\notag
\end{align}
This definition follows the usual approach to norm definition in Euclidean space.

This distance $S(t)_{(i,k)(j,l)}$ is small when $v_{i}^{k}$ and $v_{j}^{l}$ are in the same community. 
\begin{figure*}[t]
\centering
\includegraphics[width=\textwidth]{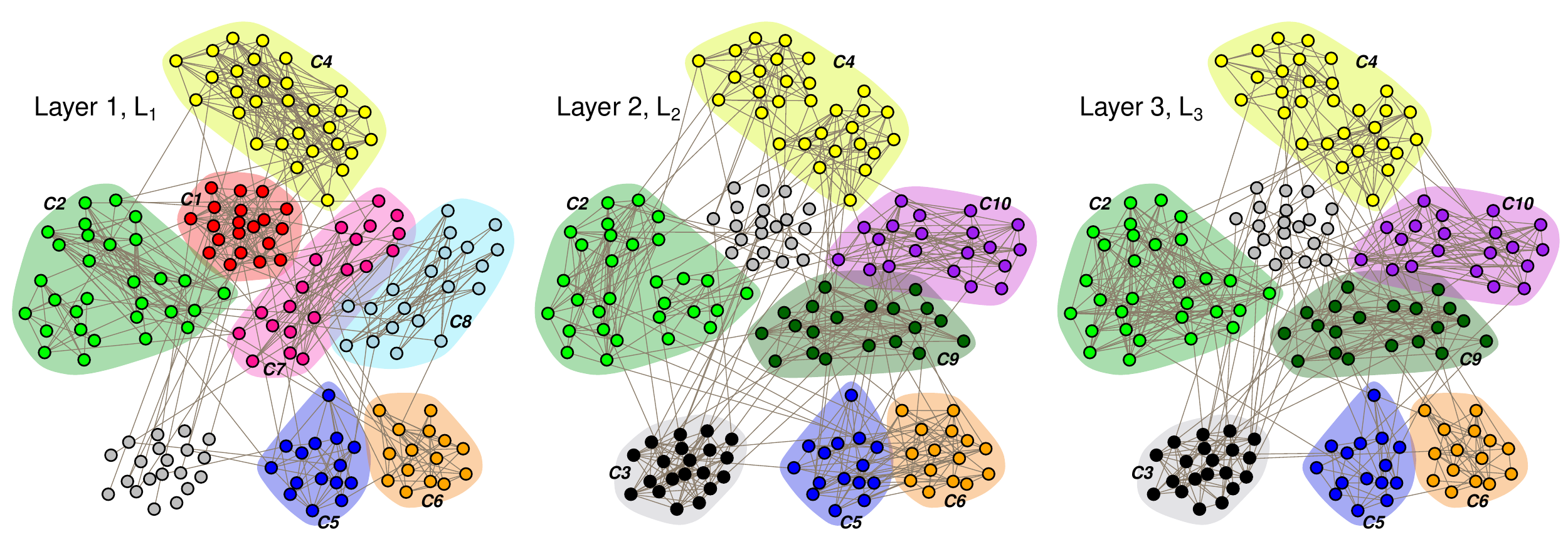}
\caption{Three-layered multiplex network: There is a one-to-one
correspondence between nodes in all three layers $L_{1}$, $L_{2}$ and $L_{3}$%
. Each community has its own node color, and community borders are indicated
by shadows which are labeled with respect to the community. Grey nodes
represent nodes with no community assignment. Community \textit{C1} (red)
exists only in $L_{1}$, while community \textit{C3} (black) is shared only
in $L_{2}$ and $L_{3}$, it does not exist in $L_{1}$. The communities 
\textit{C2} (green), \textit{C4} (yellow), \textit{C5} (blue) and \textit{C6}
(orange) are shared by all three layers. Communities \textit{C7} (pink) and 
\textit{C8} (light blue) are communities specific to $L_{1}$, but the nodes
forming these two communities regroup into two other communities \textit{C9}
(dark green) and \textit{C10} (purple) which are shared by $L_{2}$ and $L_{3}
$. See further discussion on the structure of the different communities in
Section~\protect\ref{results}.}
\label{fig:Multiplex}
\end{figure*}
The value of $s_{1}$
is small when the probabilities to reach any node in layer $L_{k}$ starting
from $v_{i}^{k}$ are approximately equal to the probabilities to reach any
node in layer $L_{l}$ starting from $v_{j}^{l}$, $\mathcal{P}%
_{(i,k)(h,k)}^{t}\simeq \mathcal{P}_{(j,l)(h,l)}^{t},\forall h=1,2,...,N$.
The value of $s_{2}$ is small when the probabilities to reach any node in
layer $L_{l}$ starting from $v_{i}^{k}$ are approximately equal to the
probabilities to reach any node in layer $L_{k}$ starting from $v_{j}^{l},$ $%
\mathcal{P}_{(i,k)(h,l)}^{t}\simeq \mathcal{P}_{(j,l)(h,k)}^{t},\forall
h=1,2,...,N$. 
We see that if $v_{i}^{k}$ and $v_{j}^{l}$ are in the same community, then both $%
s_{1}$ and $s_{2}$ are small. However, if $v_{i}^{k}$ and $v_{j}^{l}$ do not
belong to the same community, both $s_{1}$ and $s_{2}$ are large. 
The value of $s_{3}$ is small when the probabilities to reach
any node in another layer $L_{m}$, $m\neq k,l$, starting from $v_{i}^{k}$ or 
$v_{j}^{l}$ are approximately equal, $\mathcal{P}_{(i,k)(h,m)}^{t}\simeq 
\mathcal{P}_{(j,l)(h,m)}^{t},\forall h=1,2,...,|V|,\forall m\neq k,l$.

It can be verified that $S$ is symmetric, non-negative, homogeneous and
satisfies the triangle inequality. 
\subsection{Agglomerative clustering}

We use agglomerative clustering to merge nodes in communities, since such a method
allows us to use the topology of the multiplex and ensure that the obtained communities are connected. The algorithm
starts by assigning each node in each layer to its own community, and then
it iteratively merges nodes based on the average linkage criterion using the
distance matrix $S$. In order to ensure that each community, if any is
detected, is connected, we impose the criterion that only nodes and
communities having at least one within-layer or inter-layer connection
between them can be merged. Furthermore, we use the
multiplex modularity $Q_{M}$ proposed in~\cite{Mucha} to choose the best
partition as this criterion takes into account both the
within-layer and inter-layer connections of a detected community.

\section{Shared and Non-shared Communities: Illustrative Examples \label{Example}}
To provide a complete evaluation of the performance of LART, we introduce some illustrative examples meant to represent shared and non-shared
communities. Figure~\ref{fig:Multiplex} shows examples of different shared and non-shared communities and how their structures can be influenced by noise or the nature of the data used to obtain the networks. The specificity of the data can result in groups of nodes that belong to the same community although there might not be clear connectivity patterns between them. These are discussed in more detail in the following two sections. 
\subsection{Shared communities}
A shared community is a set
of nodes for which several (but not necessarily all) layers provide topological evidence that these nodes form the same shared community. An example is a set of nodes that form densely
connected communities in each of the several layers in Figure~\ref{fig:Multiplex}. The nodes that form~C3
(black) are densely connected in both layers $L_{2}$ and $L_{3}$. They do
not form a community in $L_{1}$ so~C3 is shared only in $L_{2}$ and $L_{3}$%
. A similar example are communities~C9 (dark green) and~C10 (purple) both of
which are shared by $L_{2}$ and $L_{3}$.

Detecting shared communities can help uncover hidden structures that could
otherwise go undetected when considering each layer separately. Two such
examples are communities~C2 (green) and~C4 (yellow) both of which are shared
in all three layers. Here we observe the disjoint node subsets in~C2 present in $L_{1}$ and $L_{2}$, and
similarly, disjoint subsets in~C4 present in $L_{2}$ and $L_{3}$. In such
cases, the communities might be disjoint by chance or as a result of
measurement errors.

Detecting shared community structures can also be helpful when we try to distinguish true signal
from noise.  Consider as an example communities~C5 (blue) and~C6
(orange). In $L_{1}$ and $L_{3}$, C5 and~C6 are clearly disjoint and they
are respectively shared by $L_{1}$ and~$L_{3}$. However, in~$L_{2}$ there
are high white noise levels between~C5 and~C6.
\subsection{Non-shared communities}

A non-shared community is a set of nodes which have a densely connected structural pattern specific to one layer. For example, the nodes that form community C1
(red) are densely connected in $L_{1}$, but the same nodes do not form any
communities in $L_{2}$ or $L_{3}$.

There can also exist various structural patterns between 
nodes in different layers. Therefore, same sets of nodes can form non-shared communities in one layer,
and shared communities in other layers. For example, non-shared communities~C7
(pink) and C8 (light blue) are specific for $L_{1}$. However, the union of
these nodes is the same as the union of the nodes forming~C9 and~C10 in $%
L_{2}$ and $L_{3}$.
\section{Experimental Results \label{results}}
In this section we present our experimental results based on simulated networks. We consider five different scenarios
of shared and non-shared communities in synthetic multiplexes. LART is compared to other multiplex community
detection methods: multiplex modularity maximization (MM) \cite{Mucha}, Principal Modularity
Maximization (PMM) \cite%
{Tang2009a}, and two methods for combining partitions obtained on the
separate layers using community similarity measures, topological overlap $S_{T}$ (\cite{Fortunato2010}) and Normalized Mutual Information (NMI) $S_{M}$ (\cite{Strehl2000}).
\subsection{Simulation settings}

We consider five different scenarios each one describing a different pattern of shared and non-shared communities as discussed earlier:

\subsubsection*{Scenario S1}

We consider both shared and non-shared communities, for three layers $L=3$. The motivation for this scenario are 
communities C1 and C3 with reference to Figure~\ref{fig:Multiplex}. The layers in which the
communities exist are randomly sampled, and the set of nodes forming these
communities do not form other communities in other layers. The number of
nodes on the layers is uniformly sampled from $30\leq N\leq 90$.

\subsubsection*{Scenario S2}

We consider communities shared by three layers, $L=3$. The motivation for this scenario are 
communities C2 and C4 with reference to Figure~\ref{fig:Multiplex}. For each shared community,
two layers are selected randomly. In these two layers the set of nodes
forming the community is randomly split in two or three disjoint
subsets. The number of nodes on the layers is uniformly sampled from 
$60\leq N\leq 80$.

\subsubsection*{Scenario S3}

We consider communities shared by three layers, $L=3$. The motivation for this scenario are 
communities C5 and C6 with reference to Figure~\ref{fig:Multiplex}. For two shared
communities, randomly select one layer in which, first, the within-community
edge probability of the two communities is uniformly sampled from $0.10\leq
p\leq 0.20$, and, second, white noise levels are added between the two
communities with probability uniformly sampled from $0.10\leq p\leq 0.20$.
The number of nodes on the layers is uniformly sampled from $60\leq N\leq 80$.
\begin{figure}[ptb]
\centering
\includegraphics[width=0.45\textwidth]{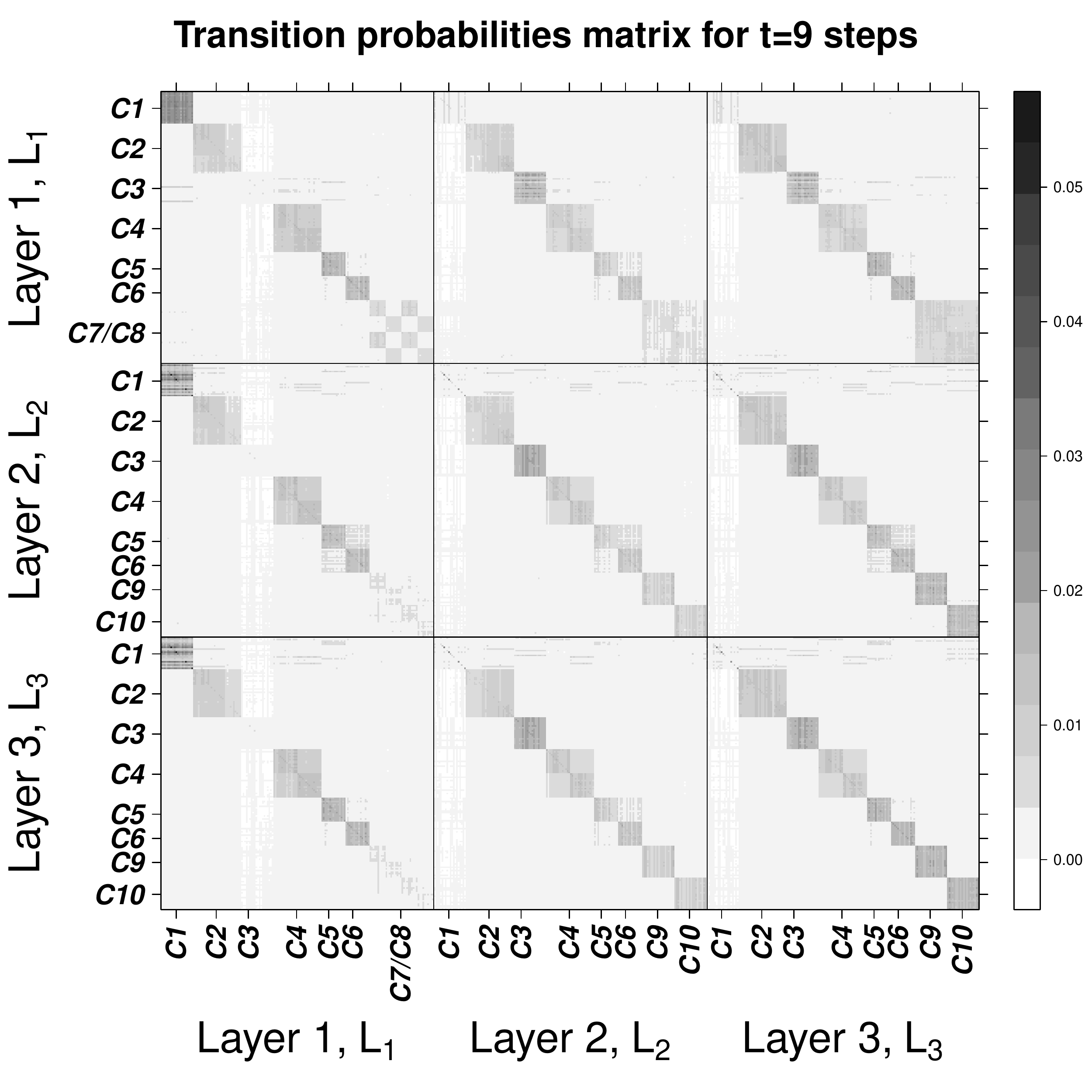}
\caption{Heat map of transition probabilities for a random walk of length $%
t=9$ steps on the multiplex network in Figure~\ref{fig:Multiplex}: 
the order of nodes in each
layer is the same, and the order is determined by the communities that nodes
belong to. In each block, the order of the transition probabilities follows
the order of communities \textit{C1, C2, C3, C4, C5, C6}; communities C7 and
C8 are in $L_{1}$, while communities C9 and C10 are in $L_{2}$ and $L_{3}.$
The probabilities to move between layers are much higher for
communities that are shared by two layers (C3). Also the probabilities to
get ``trapped" in a layer-specific community are higher than those to move
to another layer (C1 in $L_{1}$ only). }
\label{fig:HeatMap}
\end{figure}
\subsubsection*{Scenario S4}

We consider both shared and non-shared communities in three layers, $L=3$. The motivation for this scenario are 
communities C7, C8, C9 and C10 with reference to Figure~\ref{fig:Multiplex}.
Randomly select two layers
in which sets of nodes form shared communities. In the third layer, the same
set of nodes form non-shared communities. The topological structure of the
non-shared communities is simulated to resemble bipartite sets with edge
probability $p=0.4$. The number of nodes in the layers is $N=80$.
\subsubsection*{Scenario S5}

We consider both shared and non-shared communities in four layers, $L=4$, where communities can be
shared by no more than three out of the four layers. The motivation for this scenario are 
all communities of the multiplex in Figure~\ref{fig:Multiplex}. The communities are a
mixture of the patterns considered in Scenarios $1$ to $4$. The number of nodes on
the layers is sampled from $150\leq N\leq 180$.
\subsection{Comparative performance}
\begin{table*}[t]
\centering
\caption{Performance of competing algorithms in five simulated scenarios }
\label{nmires}\centering
\renewcommand{\arraystretch}{1.3}
\scalebox{0.9}{
\begin{tabular}
[c]{||l||l|l|l|l|l||l|l|l|l|l|l||}\hline
&\multicolumn{5}{c||}{NMI index (mean$\pm$std.dev.)}&\multicolumn{5}{|c|}{FM index (mean$\pm$std.dev.)}\\\hline
 & \multicolumn{5}{c||}{Simulation Scenario}  & \multicolumn{5}{|c|}{Simulation Scenario}\\\hline 
Algorithm& \textit{S1} & \textit{S2} & \textit{S3} & \textit{S4} & \textit{S5}& \textit{S1} & \textit{S2} & \textit{S3} & \textit{S4} & \textit{S5}\\\hline
\textit{LART} & $0.85\pm0.05$ & $0.95\pm0.05$ & $0.94\pm0.07$ & $0.84\pm0.07$ & $0.88\pm0.08$ & $0.80\pm0.12$ & $0.97\pm0.04$ & $0.97\pm0.05$ & $0.78\pm0.12$ & $0.80\pm0.14$\\\hline
\textit{$Fixed1$} & $0.70\pm0.11$ & $0.86\pm0.10$ & $0.88\pm0.13$ & $0.79\pm0.03$ & $0.68\pm0.02$ & $0.72\pm0.20$ & $0.91\pm0.11$ & $0.87\pm0.10$ & $0.70\pm0.01$& $0.57\pm0.02$\\\hline
\textit{MM} &  $0.80\pm0.07$ &  $0.90\pm0.07$ &  $0.79\pm0.11$ & $0.73\pm0.04$ & $0.82\pm0.07$&  $0.70\pm0.19$ &  $0.93\pm0.06$ &  $0.85\pm0.1$ & $0.71\pm0.03$ & $0.75\pm0.11$\\\hline
\textit{$Fixed2$} & $0.64\pm0.14$ &  $0.97\pm0.05$ &  $0.88\pm0.11$ & $0.68\pm0.02$ & $0.66\pm0.02$&  $0.55\pm0.30$ &  $0.92\pm0.10$ &  $0.97\pm0.06$ & $0.65\pm0.03$ & $0.57\pm0.03$\\\hline
\textit{$S_{T}$} & $0.78\pm0.07$ & $0.70\pm0.05$ & $0.68\pm0.08$ & $0.77\pm0.06$ & $0.75\pm0.08$& $0.72\pm0.17$ & $0.78\pm0.07$ & $0.72\pm0.06$ & $0.61\pm0.06$ & $0.65\pm0.12$ \\\hline
\textit{$S_{M}$} & $0.78\pm0.01$ & $0.70\pm0.05$ & $0.68\pm0.09$ & $0.77\pm0.05$& $0.75\pm0.08$ & $0.72\pm0.17$ & $0.78\pm0.09$ & $0.72\pm0.06$ & $0.61\pm0.08$ & $0.66\pm0.11$ \\\hline
\textit{PMM} & $0.62\pm0.13$ & $0.97\pm0.04$ & $0.99\pm0.02$ & $0.69\pm0.01$ & $0.73\pm0.16$& $0.56\pm0.23$ & $0.97\pm0.04$ & $0.99\pm0.01$ & $0.66\pm0.01$& $0.66\pm0.21$\\\hline
\end{tabular}}
\end{table*}
Each community (or indicated disjoint subsets of nodes in a community) has a uniformly sampled
within-community edge probability, $0.25\leq p\leq 0.40$ (unless explicitly
stated otherwise). For each one of the five scenarios, we
randomly generate $100$ synthetic multiplexes. Each multiplex is simulated using the following steps: first, we randomly sample the number of nodes and communities in the multiplex; second, each community is assigned the same set of nodes in the different layers; third, in each layer the edges between the nodes in these sets are simulated according to the community structure of the respective scenario. Last but not least, on each layer noise is added to represent the random connections between the communities.

The multiplexes generated in each simulation were analyzed using the LART
algorithm. For each multiplex, we select the length of the random walk $t$
using the rule of thumb proposed in~\cite{Pons} which states that for dense networks $t=3$ is sufficient to explore the local topology of the network. Since we work with multiplex networks, we consider $t=3L$ where $L$ is the number of layers in the multiplex. In this way, we allow enough steps for the random walker to explore the local topology of each node in all layers. We provide an illustrative example of the transition probabilities for a short random walk of length $t=9$ on a three-layered multiplex network, Figure~\ref{fig:HeatMap}. This figure shows that multiplex random walks of length $t=3L$ capture the local community structures of the multiplex. We fix $\varepsilon=1$ which is equivalent to adding a self-loop to each node in every layer. 

Using the same simulated data, we also test the MM and the PMM algorithms. For MM, we use the supra-adjacency
matrix $\mathcal{A}^{*}$ as input. The multiplex modularity $Q_{M}$ used by
both~LART and~MM needs the specification of a resolution parameter $\gamma$.
In our application, we consider $\gamma=1$. PMM is designed to find a shared
community structure for all layers. Since PMM uses the $k$-means algorithm
to merge communities, we obtain results for different values of centers $%
k=1,2,...10$, and record the best ones only. Since LART reduces to WalkTrap for $L=1$ (with the exception of the linkage criterion), we add two
methods where best partitions are initially identified separately in each
layer using WalkTrap. Then the similarity between
communities in the different layers is assigned using either $S_{T}$, which is based on the
relative overlap between communities, or $%
S_{M}$, which is based on the NMI between
two communities. Any clustering method can be used to merge communities between layers with respect to the resulting similarities, and in this work we use affinity propagation \cite{Frey2007} since it does not require a priori knowledge about the number of clusters. Finally, we
consider each multiplex network, but with fixed weights, $%
\omega_{i;kl}=\omega $ $\forall i,k,l$. We use both LART and MM to detect communities in such a
setting, and present results for $\omega=1$ since it provides good comparative results for detecting communities shared by several layers.  
We annotate these two applications with $Fixed1$ for the LART framework and $Fixed2$ for the MM framework.

The partitions obtained from the competing algorithms are assessed using two
different relative performance measures: the generalized Fowlkes-Mallows
Index (FM)~\cite%
{Fowlkes2012}, and the NMI measure~\cite{Strehl2000}. Both measures take values between $0$ and $1$,
and they are equal to $1$ when the true communities are correctly
identified. The comparative analysis results for NMI and FM are summarized
in Table~\ref{nmires}. For each of the 5
scenarios, the average results ($\pm$ std. dev.) over the $100$ simulated
multiplex networks are presented for each of the algorithms. We use the two-sided Kolmogorov-Smirnov statistic to test against the null hypothesis that the results obtained from two different methods come from the same distribution. If there is enough evidence to reject the null hypothesis, we assume that the difference between two methods is statistically significant.

The relative performance of
LART when detecting communities shared across all layers is very
competitive. Scenarios S2 and S3 are cases in which the communities are
shared across all layers. These scenarios are favorable for methods that
find communities shared across all layers and are designed to be robust to
noise. The results suggest this is the case since PMM performs slightly better relative
to the other methods. This is true when an appropriate number of center $k$
has been selected. $Fixed2$ also performs better than LART for S3 since the
MM framework for fixed weights is well designed to detect
communities shared across all layers. Furthermore, the performance of
methods that combine partitions obtained on the separate layers, $S_{T}$ and 
$S_{M}$, is poor compared to the other methods.

Scenarios S1, S4 and S5 show the main strength of LART. LART is better
able to detect layer specific communities and communities that are shared
across several but not all layers. Furthermore, the results from S4 show that the method
distinguishes between different topological structures of communities in
different layers. The weaker performance for $Fixed1$ and $Fixed2$ show the
gains introduced by locally adapting inter-layer weights $\omega_{i;k,l}$. We produce additional simulations for $\omega=0.5$ and $%
\omega=0.1$ which are not included here. The obtained results for $\omega=0.5,0.1$ are lower than the presented ones for $\omega=1$. The only exception is S4 for which weaker couplings
$\omega=0.5,0.1$ show better performance than $\omega=1$ for the MM framework.

Additionally, we obtain results for varying parameter $\gamma=0.25,0.5,0.75,1,1.25,1.5,1.75,2,2.25,2.5$
in the case of LART and MM (and $\omega=1,0.5,0.1$). Results not included here show that there exist values of $\gamma\neq1$ for which the performance of both LART and MM improves with respect to their performance for $\gamma=1$. However, even when the best results are selected and compared, the performance of LART for S1, S4 and S5 is higher than the performance of MM. In addition, there exist $\gamma$ values for which MM with $\omega=1$ and $\omega=0.5$ performs without error for S2 and S3.

\section{Conclusions\label{conc}}

We distinguish between shared and non-shared community structures on a
multiplex, and propose the LART algorithm which is designed to detect both
types of communities. The algorithm takes advantage of the complex
multiplex structure, and adapts the transition probabilities of the random
walk to depend on the topological similarity between layers at any given
node. 

One advantage of LART is that it requires the definition of only one
parameter $t$ which determines the length of the random walk. The value of $t$ can vary within 
some boundaries as long as the random walks are short enough to explore only the local community structure. Therefore, future
work is required to adopt an exact way of choosing a range of suitable values for $t$. 

Even so, LART performs very well in detecting communities shared by a subset of layers,
and it is competitive to methods that detect communities shared by all
layers. Future work would include comparison to other methods mentioned in the review section. The method will be further implemented to real world systems to showcase the benefits of the LART algorithm.
\IEEEtriggeratref{12}
\bibliographystyle{IEEEtran}
\bibliography{IEEEabrv,PaperLib}

\end{document}